\documentclass[aps,prd,reprint,floatfix,amsmath,amssymb,superscriptaddress,notitlepage,10pt]{revtex4-2}

\usepackage{hyperref}
\usepackage{bm}
\usepackage{graphicx}
\usepackage{xcolor}
\usepackage{comment}

\newcommand{\Mpc}{\mathrm{Mpc}}
\newcommand{\hMpc}{h^{-1} \, \mathrm{Mpc}}
\newcommand{\Msun}{\mathrm{M}_\odot}

\newcommand{\mnras}{Mon. Not. Roy. Astron. Soc.}
\newcommand{\apjl}{Astrophys. J. Lett.}
\newcommand{\apjs}{Astrophys. J. Suppl.}
\newcommand{\jcap}{J. Cosmol. Astropart. Phys.}
\newcommand{\physrep}{Phys. Rept.}
\newcommand{\aap}{Astron. Astrophys.}

\newcommand{\araa}{Annu. Rev. Astron. Astrophys.}
\newcommand{\pasj}{Publ. Astron. Soc. Jpn.}
\newcommand{\apss}{Astrophys. Space Sci.}
\newcommand{\procspie}{SPIE}

\begin{document}

\title{Super-sample covariance of the thermal Sunyaev--Zel'dovich effect}

\author{Ken Osato}
\email[]{ken.osato@iap.fr}
\affiliation{Institut d'Astrophysique de Paris, Sorbonne Universit\'e, CNRS, UMR 7095,
98bis boulevard Arago, 75014 Paris, France}

\author{Masahiro Takada}
\affiliation{Kavli Institute for the Physics and Mathematics of the Universe (WPI),
The University of Tokyo Institutes for Advanced Study,
The University of Tokyo, 5-1-5 Kashiwanoha, Kashiwa, 277-8583 Chiba, Japan}

\begin{abstract}
The thermal Sunyaev--Zel'dovich (tSZ) effect is a powerful probe of cosmology.
The statistical errors in the tSZ power spectrum measurements are dominated
by the presence of massive clusters in a survey volume that
are easy to identify on individual cluster basis.
First, we study the impact of super sample covariance (SSC)
on the tSZ power spectrum measurements, and find that the
sample variance is dominated by the connected non-Gaussian (cNG) covariance
arising mainly from Poisson number fluctuations of
massive clusters in the survey volume.
Second, we find that removing such individually-detected,
massive clusters from the analysis significantly reduces the cNG contribution,
thereby leading the SSC to be a leading source of the sample variance.
We then show, based on Fisher analysis, that the power spectrum measured
from the remaining diffuse tSZ effects can be used to obtain tight constraints
on cosmological parameters as well as the hydrostatic mass bias parameter.
Our method offers complementary use of individual tSZ cluster counts
and the power spectrum measurements of diffuse tSZ signals for
cosmology and intracluster gas physics.
\end{abstract}

\date{\today}

\maketitle
\section{Introduction}
\label{sec:introduction}

The anisotropy of cosmic microwave background (CMB) is one of
the most important probes of physical states in the Universe.
Among different components of the CMB anisotropy,
the Sunyaev--Zel'dovich (SZ) effect \cite{Sunyaev1970,Sunyaev1972,Sunyaev1980} provides us
with rich information of cosmic structures of the late-time Universe
(for reviews, see Refs.~\cite{Carlstrom2002,Kitayama2014}).
The SZ effect is divided into two classes according to physical processes
giving rise to the anisotropy.
One is the thermal SZ effect (tSZ), where energy is transferred from
hot electrons to CMB photons through inverse Compton scattering.
The other one is the kinetic SZ effect (kSZ), which is caused by the Doppler effect due to
bulk motion of electrons with respect to the rest frame of CMB photons.
The tSZ effect arises mainly from thermal electrons,
most of which reside in galaxy clusters,
and enable one to detect distant galaxy clusters up to high redshifts because
the cosmological surface brightness dimming is compensated by
the increase of CMB photon number density at higher redshifts.
Since the abundance and spatial distribution
of galaxy clusters reflect the degree of
the structure formation of the Universe,
we can take advantage of SZ effects to constrain cosmological parameters
\cite{Haiman2001,Komatsu2002,Takada2007,Oguri2011,Salvati2018}.
Since the amplitude of the tSZ effect is larger than the kSZ effect,
and the tSZ effect is easier to be distinguished from other anisotropies
due to specific frequency dependence,
the tSZ effect has now been intensively studied.

The tSZ effect has already been measured by several survey programs.
The \textit{Planck} satellite has conducted
the measurement for all-sky coverage \cite{Planck2015XXII,Planck2015XXIV},
and ground-based telescopes aim to measure the tSZ effect
with high angular resolution ($\sim 1 \, \mathrm{arcmin}$), which includes
Atacama Cosmology Telescope (ACT) \cite{Swetz2011,Dunkley2011},
South Pole Telescope (SPT) \cite{Benson2014,George2015},
Simons Observatory \cite{Ade2019},
and CMB-S4 \cite{Abazajian2016}.

As a summary statistics of the tSZ effect, power spectrum is commonly employed
because the theoretical approach is readily available \citep{Hill2013}.
Among them, the halo model prescription \cite{Komatsu1999,Cooray2002,Shaw2010,Trac2011}
can reproduce the results of hydrodynamical simulations \cite{McCarthy2014,Dolag2016} and
is widely employed in practical analyses \cite{Planck2015XXII,Bolliet2018}.
In addition, by cross-correlating the tSZ map with
other observables of the large-scale structure,
e.g., cosmic shear or nearby galaxies, the significance of the measurements can be enhanced
\cite{Fang2012,VanWaerbeke2014,Battaglia2015,Hojjati2017,Osato2018,Makiya2018,Osato2020,Makiya2020}.
Since the tSZ effect is sourced by hot free electrons and
most of them are found in galaxy clusters, i.e., massive halos,
the halo model picture is justified.
Similarly, we can also estimate the covariance matrix based on the halo model prescription.
However, the covariance matrix suffers from
large-scale mass fluctuations with wavelengths comparable with or greater than
a size of the survey volume, which is not a direct observable.
That additional contribution in the covariance is referred to
as the super-sample covariance (SSC) \cite{Takada2013}.
If the SSC contribution is not taken into account,
the covariance matrix is underestimated and
the statistical significance is overestimated.
In order to carry out the analysis with the power spectrum in an unbiased way,
we need to incorporate such additional contributions.

In Refs~\cite{Horowitz2017,Bolliet2018}, it is shown that
the connected non-Gaussian term is a dominant contribution
in the covariance of the power spectrum.
Furthermore, this term is sourced mainly by massive clusters  at low redshifts.
Therefore, removing such clusters from the measurement by masking the regions
leads to suppression of the covariance matrix and enhancement of the significance
though some fraction of signals can be missed \cite{Shaw2009}.
In this paper, we estimate the SSC contribution
for the tSZ effect based on the halo model prescription
and quantify the effects on cosmological
parameter estimation with the tSZ power spectrum.
Furthermore, we also investigate the effect when massive clusters are masked.

This paper is organized as follows.
First, we review the halo model prescription to compute
the tSZ power spectrum in Section~\ref{sec:power_spectrum} and
the covariance matrix in Section~\ref{sec:covariance}.
In Section~\ref{sec:experiment}, we give specific details on experimental conditions.
In Section~\ref{sec:results}, we present results of statistical significance and
forecasts of constraints on parameters.
We make concluding remarks in Section~\ref{sec:conclusions}.

Throughout this paper, we assume
the flat $\Lambda$ cold dark matter (CDM) Universe.
Otherwise stated, we adopt the best fit cosmological parameters inferred with
observations of temperature and polarization anisotropies
and gravitational lensing of CMB (TT,TE,EE+lowE+lensing)
measured by \textit{Planck} \citep{Planck2018I,Planck2018VI}:
the physical CDM density $\Omega_\mathrm{c} h^2 = 0.12011$,
the physical baryon density $\Omega_\mathrm{b} h^2 = 0.022383$,
the acoustic angular scale $100 \theta_* = 1.041085$,
the amplitude and slope of the scalar perturbation $\ln (10^{10} A_\mathrm{s}) = 3.0448$,
$n_\mathrm{s} = 0.96605$ at the pivot scale $k_\mathrm{piv} = 0.05 \, \Mpc^{-1}$, and
the optical depth $\tau_\mathrm{reio} = 0.0543$.
The neutrino component is composed of two massless neutrinos and one massive neutrino
with the mass $m_\nu = 0.06 \, \mathrm{eV}$,
and the massive neutrino density parameter is
$\Omega_\nu h^2 = m_\nu / (93.14 \, \mathrm{eV}) = 0.000644$.
Thus, the matter component is composed of CDM, baryons, and massive neutrinos,
and the matter density parameter is
$\Omega_\mathrm{m} = \Omega_\mathrm{c} + \Omega_\mathrm{b} + \Omega_\nu = 0.31816$.
Then, the energy density of cosmological constant is given as
$\Omega_\Lambda = 1 - \Omega_\mathrm{m} = 0.68184$.
The amplitude of matter fluctuation at the scale of $8 \, \hMpc$ is
$\sigma_8 = 0.811$.
For notation, the symbol ``$\log$'' denotes common logarithm ``$\log_{10}$''
and the symbol ``$\ln$'' denotes natural logarithm ``$\log_e$''.

\section{The tSZ Power Spectrum}
\label{sec:power_spectrum}

\subsection{Halo Model of the tSZ Power Spectrum}
Here, we briefly review the halo model calculations of
the tSZ power spectrum.
The temperature variation $\Delta T$ due to
the tSZ effect \cite{Sunyaev1970,Sunyaev1972,Sunyaev1980} is given by
\begin{equation}
\frac{\Delta T}{T_\mathrm{CMB}} = g_\nu(x) y =
g_\nu (x) \frac{\sigma_\mathrm{T}}{m_\mathrm{e} c^2} \int
P_\mathrm{e} \, \mathrm{d}l ,
\end{equation}
where $y$ is the Compton-$y$ parameter,
$T_\mathrm{CMB} = 2.7255 \, \mathrm{K}$ is the CMB temperature,
$\sigma_\mathrm{T}$ is the Thomson scattering cross-section,
$m_\mathrm{e}$ is the electron mass,
$c$ is the speed of light,
$P_\mathrm{e}$ is the free electron pressure,
and in the last term,
the line-of-sight integration is carried out with respect to
the physical length $l$.
The frequency dependent part $g_\nu (x)$ is given by
\begin{equation}
g_\nu (x) = x \frac{e^x-1}{e^x+1} - 4 , \ x \equiv \frac{h \nu}{k_\mathrm{B} T_\mathrm{CMB}},
\end{equation}
where $\nu$ is the frequency,
$h$ is the Planck constant, and
$k_\mathrm{B}$ is the Boltzmann constant.

First, we give formulations for halo model calculation of
the tSZ power spectrum \cite{Komatsu1999,Cooray2002}:
\begin{align}
C (\ell) =& C^\mathrm{1h} (\ell) + C^\mathrm{2h} (\ell), \\
C^\mathrm{1h} (\ell) =& \int_0^{z_\mathrm{reio}} \mathrm{d}\chi \frac{\mathrm{d}^2 V}{\mathrm{d}\chi \mathrm{d}\Omega} \nonumber \\
& \times \int_{M_\mathrm{min}}^{M_\mathrm{max}} \!\! \mathrm{d}M
\label{eq:hm_1h}
\frac{\mathrm{d}n_\mathrm{h}}{\mathrm{d}M} (M, z) |\tilde{y} (\ell; M, \chi)|^2 , \\
\label{eq:hm_2h}
C^\mathrm{2h} (\ell) =& \int_0^{z_\mathrm{reio}} \mathrm{d}\chi \frac{\mathrm{d}^2 V}{\mathrm{d}\chi \mathrm{d}\Omega}
P_\mathrm{L} \left( k = \frac{\ell + 1/2}{\chi} ; \chi \right) \nonumber \\
& \times
\left[ \int_{M_\mathrm{min}}^{M_\mathrm{max}} \!\! \mathrm{d}M \frac{\mathrm{d}n_\mathrm{h}}{\mathrm{d}M} (M, z)
b_\mathrm{h}(M, z) \tilde{y} (\ell; M, \chi) \right]^2 ,
\end{align}
where $\chi$ is comoving distance \footnote{The comoving distance also serve as
the indicator of the cosmic time.},
$\mathrm{d}^2V/\mathrm{d}\chi \mathrm{d}\Omega = \chi^2$ is the comoving volume
per unit comoving distance and unit solid angle,
$\mathrm{d}n_\mathrm{h}/\mathrm{d}M (M, z)$ is the halo mass function,
$P_\mathrm{L}(k; \chi)$ is the linear matter power spectrum
at redshift $z$ corresponding to the comoving radial distance $\chi(z)$,
and $b_\mathrm{h}(M, z)$ is the halo bias.
The quantity $\tilde{y} (\ell; M, \chi)$ is the Fourier transform
of the electron pressure profile. The modeling of the pressure profile of
halos will be discussed later in this section.
The one-halo term (Eq.~\ref{eq:hm_1h}) corresponds to the correlation
between the same halo and the two-halo term (Eq.~\ref{eq:hm_2h})
denotes the one between different halos.
We assume the reionization occurs instantaneously at $z_\mathrm{reio} = 7$
and at this redshift, all hydrogen and helium are fully ionized.
We use the linear Boltzmann code \texttt{CLASS} \cite{Blas2011}
to compute the linear matter power spectrum $P_\mathrm{L} (k, z)$.
In the mass integration, we adopt the virial mass $M_\mathrm{vir}$ as the halo mass definition
and employ $M_\mathrm{min} = 10^{11}\, h^{-1} \, \Msun$,
$M_\mathrm{max} = 10^{16}\, h^{-1} \, \Msun$
for the lower and upper limits of the halo mass integral, respectively.
The virial radius $R_\mathrm{vir}$ for the halo with the virial mass $M_\mathrm{vir}$
is related as follows:
\begin{equation}
M_\mathrm{vir} = \frac{4\pi}{3} \Delta_\mathrm{vir} (z) \rho_\mathrm{cr}(z) R_\mathrm{vir}^3 ,
\end{equation}
where $\rho_\mathrm{cr} (z)$ is the critical density.
The virial overdensity $\Delta_\mathrm{vir}$ is given as \cite{Bryan1998}
\begin{equation}
\Delta_\mathrm{vir} = 18 \pi^2 + 82 (\Omega_\mathrm{m} (z) - 1) -
39 (\Omega_\mathrm{m} (z) - 1)^2 ,
\end{equation}
where
\begin{equation}
\Omega_\mathrm{m} (z) = \Omega_\mathrm{m} (1+z)^3 E^{-2} (z) ,
\end{equation}
and the expansion factor is defined as
\begin{equation}
E (z) = \frac{H(z)}{H_0} = [\Omega_\mathrm{m}(1+z)^3 + \Omega_\Lambda ]^{\frac{1}{2}} .
\end{equation}
Since fitting formulas of halo mass function, halo bias, and electron pressure profile
adopt halo mass definitions different from the virial mass,
we also use alternative halo mass definitions, $M_\mathrm{500}$ and $M_\mathrm{200b}$:
\begin{align}
M_\mathrm{500} &= \frac{4\pi}{3} 500 \rho_\mathrm{cr}(z) R_\mathrm{500}^3 , \\
M_\mathrm{200b} &= \frac{4\pi}{3} 200 \rho_\mathrm{m}(z) R_\mathrm{200b}^3 ,
\end{align}
where $\rho_\mathrm{m} (z) = \rho_\mathrm{cr}(z) \Omega_\mathrm{m} (z)$.
These halo masses can be converted from the virial mass
if we assume that the density profile of halos follow
Navarro--Frenk--White profile \cite{Navarro1996,Navarro1997}:
\begin{equation}
\rho (r) = \frac{\rho_\mathrm{s}}{(r/r_\mathrm{s}) (1+r/r_\mathrm{s})^2} .
\end{equation}
The scale density $\rho_\mathrm{s}$ is determined by the relation:
\begin{equation}
M_\mathrm{vir} = \int_0^{R_\mathrm{vir}} \!\! \mathrm{d}r \, 4\pi r^2 \rho(r) ,
\end{equation}
and the scale radius $r_\mathrm{s}$ is determined
from the mass-concentration relation:
\begin{equation}
c_\mathrm{vir}(M_\mathrm{vir}, z) = \frac{R_\mathrm{vir}}{r_\mathrm{s}},
\end{equation}
where we adopt the fitting formula calibrated by $N$-body simulations in Ref.~\cite{Klypin2016}.
In addition, for halo mass function and halo bias,
we adopt the fitting formulas calibrated by $N$-body simulations:
the fitted halo mass function with respect to $M_\mathrm{200b}$ in Ref.~\cite{Bocquet2016} and
the fitted halo bias in Ref.~\cite{Tinker2010}.

Next, we derive the expression for the Fourier transform of the Compton-$y$ from a single halo
$\tilde{y} (\ell; M, z)$:
\begin{align}
\tilde{y} (\ell; M, \chi) = & \frac{4\pi R_s}{\ell_s^2} \frac{\sigma_\mathrm{T}}{m_\mathrm{e} c^2}
\nonumber \\
& \times \int \mathrm{d}x \, x^2 P_\mathrm{e} (x)
\frac{\sin ((\ell + 1/2) x/\ell_s)}{(\ell + 1/2) x/\ell_s},
\end{align}
where $x = r/R_s$, $\ell_s = D_A (\chi) /R_s$,
$R_s$ is the arbitrary scale radius,
$D_A (\chi)$ is the angular diameter distance.
For pressure profile of free electron,
we make use of the universal pressure profile \cite{Nagai2007} calibrated
by \textit{Planck} observations \cite{PlanckIntermediateV}:
\begin{align}
\frac{P_\mathrm{e} (r)}{P_{500}} =& p(x) \left[
\frac{M^{\mathrm{HSE}}_{500}}{3 \times 10^{14} \, h_{70}^{-1} \, \Msun}
\right]^{0.12} , \\
p(x) \equiv& \frac{P_0}{(c_{500} x)^\gamma [1 + (c_{500} x)^\alpha ]^{(\beta - \gamma)/\alpha}} , \\
P_{500} =& 1.65 \times 10^{-3} E(z)^{\frac{8}{3}} \nonumber \\
& \times \left[ \frac{M^{\mathrm{HSE}}_{500}}{3 \times 10^{14} \, h_{70}^{-1} \, \Msun}
\right]^{\frac{2}{3}} h_{70}^2 \, \mathrm{keV} \, \mathrm{cm}^{-3} ,
\end{align}
where $(P_0, c_{500}, \gamma, \alpha, \beta) = (6.41, 1.81, 0.31, 1.33, 4.13)$,
$x = r/R^\mathrm{HSE}_{500}$, and $h_{70} \equiv  h / 0.7$.
In the calibration of the pressure profile,
the mass is determined with hydrostatic equilibrium assumption,
where only thermal pressure is balanced with the self-gravity of the halo.
However, in addition to thermal pressure, non-thermal processes, e.g., turbulence or magnetic field,
could also contribute the total pressure supporting
the mass of galaxy clusters \cite{Suto2013,Nelson2014,Shi2015,Vazza2018}.
Thus, the hydrostatic mass $M^\mathrm{HSE}_{500}$ is generally
lower than the true mass $M_\mathrm{500}$.
In order to take this effect into account, we parametrize the mass and radius
with the hydrostatic mass bias parameter $b_\mathrm{HSE}$:
\begin{align}
M^\mathrm{HSE}_{500} &= (1-b_\mathrm{HSE}) M_{500} , \\
R^\mathrm{HSE}_{500} &= (1-b_\mathrm{HSE})^{\frac{1}{3}} R_{500} .
\end{align}
We adopt the fiducial value $b_\mathrm{HSE} = 0.2$, which is suggested
by mass calibration measurements \cite{Medezinski2018,Miyatake2019}
and hydrodynamical simulations \cite{Lau2013}.
The amplitude of the tSZ power spectrum is sensitive to
the hydrostatic bias parameter and thus,
the parameter can be constrained through the power spectrum
in a way complementary to the mass calibration measurements.

\subsection{Selection Function of Massive Clusters}
\begin{figure}[htbp]
  \centering
  \includegraphics[width=\columnwidth,clip]{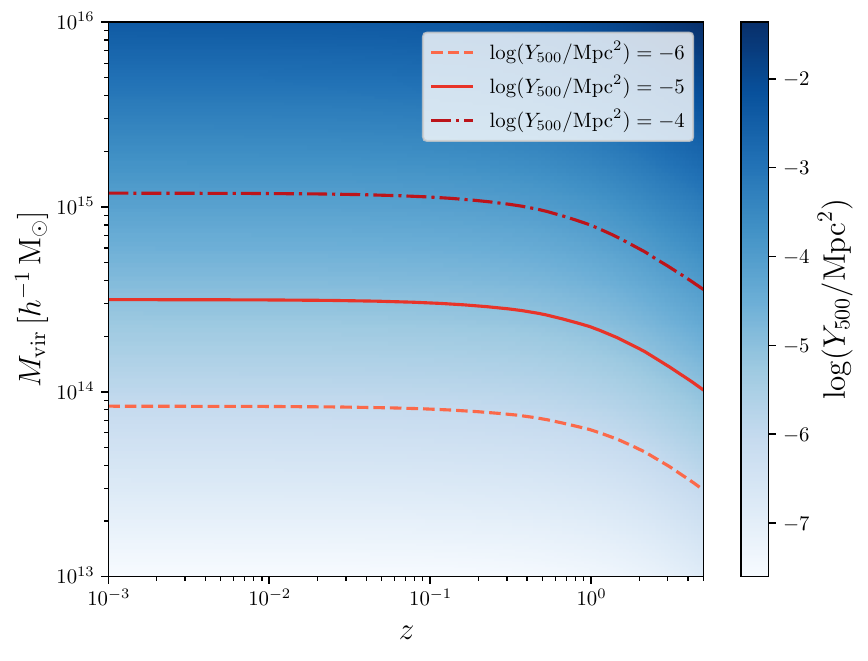}
  \caption{The integrated Compton-$y$ parameter
  for clusters of given redshift $z$ and virial mass $M_\mathrm{vir}$.
  The dashed, solid, and dot-dashed lines correspond to $\log (Y_{500}/\mathrm{Mpc}^2)
  = -6, -5, -4$, respectively.
  For each threshold, clusters located at upper regions divided by the line
  are masked.}
  \label{fig:Y}
\end{figure}

\begin{figure}[htbp]
  \centering
  \includegraphics[width=\columnwidth,clip]{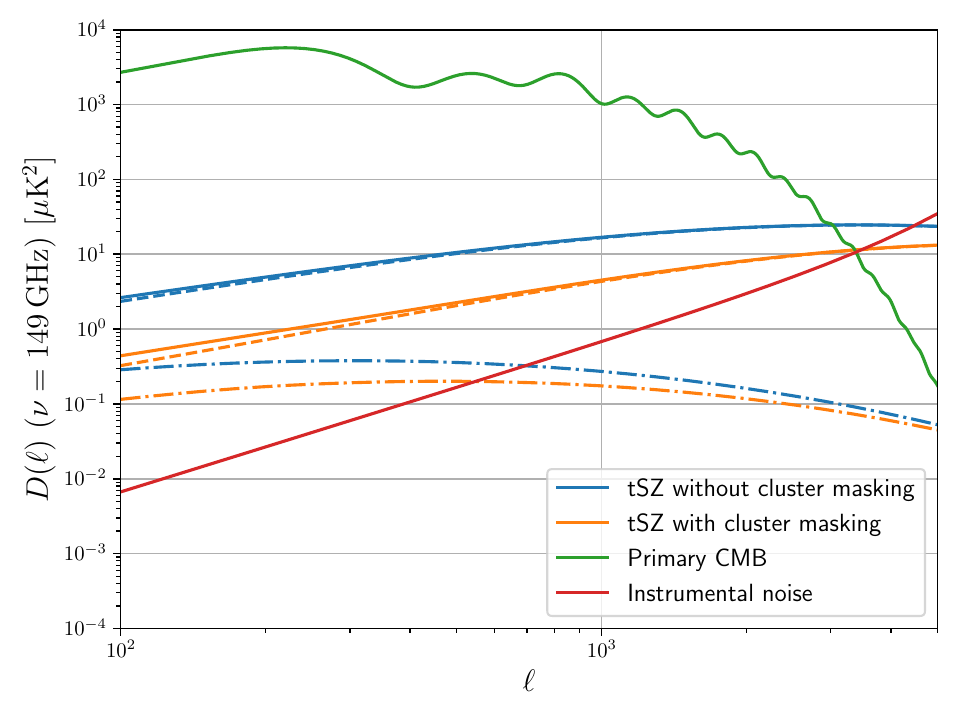}
  \caption{The power spectra of temperature anisotropy due to the tSZ effect
  with and without cluster masking. For comparison,
  the primary CMB anisotropy power spectrum
  and instrumental noise power spectrum (see Section~\ref{sec:experiment}) are also shown.
  For halo model calculations, the dashed (dot-dashed) lines correspond to
  one-halo (two-halo) term contribution.
  Instead of the raw power spectrum $C (\ell)$,
  we show $D(\ell) = \ell (\ell + 1)/ (2 \pi) C (\ell)$
  at the frequency $\nu = 149\,\mathrm{GHz}$.
  For cluster masking,
  we adopt the threshold $\log (Y_\mathrm{thres}/\mathrm{Mpc}^2) = -5$.}
  \label{fig:cl}
\end{figure}

Here, we introduce an observable in cluster survey relevant for the tSZ effect.
For each cluster, the integrated flux of tSZ effect corresponds to
the thermal energy stored in the galaxy cluster.
We define the three-dimensional integrated Compton-$y$ parameter,
which is denoted as $Y_{500}$:
\begin{equation}
Y_{500} (M, z) = \frac{\sigma_\mathrm{T}}{m_\mathrm{e} c^2}
\int_0^{R_{500}} \mathrm{d}r \, 4 \pi r^2 P_\mathrm{e}(r; M, z) .
\end{equation}
This quantity is proportional to the thermal energy of gas in the galaxy cluster
and can be measured in the tSZ survey
if the redshift of the cluster is known
\footnote{In some literature,
another definition $Y'_{500} = Y_{500}/D_A^2$ is employed.
In this definition, the unit is normally $\mathrm{arcmin}^2$ and
the quantity is computed by integrating observed SZ flux in angular space.
For detailed discussions, see Section~3.1 in Ref.~\cite{PlanckIntermediateV}.}.
Furthermore, this quantity exhibits a tight scaling relation
with the halo mass \cite{Angulo2012}.
As we will show later, since the dominant source of the tSZ covariance
is caused by massive clusters, we study how masking of such massive clusters,
based on the integrated Compton-$y$ parameter, helps reduce the sample covariance.
We introduce the selection function $S(M, z)$
based on the integrated Compton-$y$ parameter:
\begin{equation}
\label{eq:selection}
S (M, z) =
\begin{cases}
  1 & (Y_{500} (M, z) \leq Y_\mathrm{thres}) \\
  0 & (Y_{500} (M, z) > Y_\mathrm{thres}),
\end{cases}
\end{equation}
where $Y_\mathrm{thres}$ is the threshold value.
When cluster masking is applied,
this selection function is inserted in the mass and redshift integrations
in the halo model expressions (see Section~\ref{sec:masking_hm}).
In Figure~\ref{fig:Y}, we show how
the integrated Compton-$y$ parameter varies with
the virial mass and redshift, and the different lines denote three representative values of
$\log (Y_{500}/\mathrm{Mpc}^2) = -6, -5, -4$.
In Figure~\ref{fig:cl}, we show the tSZ power spectrum along with CMB primary spectrum
and noise power spectrum, which will be discussed in Section~\ref{sec:experiment}.
The tSZ power spectra with and without cluster masking are shown.
When cluster masking is applied, the amplitude is suppressed by less than half
but the covariance matrix has also been reduced as shown in Section~\ref{sec:covariance}.
Both of one-halo (dashed lines) and two-halo (dot-dashed lines) terms are reduced by cluster masking,
but the one-halo term is more suppressed because the main source of the term is
nearby massive clusters.
As a result, the signal-to-noise ratio is enhanced by masking clusters.
We will discuss the statistical significance in Section~\ref{sec:SN}.

\section{Covariance of the tSZ Power Spectrum}
\label{sec:covariance}

\subsection{Compton-$y$ with the Survey Mask}
Here, we derive formulas of the covariance matrix
of the Compton-$y$ power spectrum with taking into account the survey mask.
First, we define the survey window function $W (\bm{\theta})$
and the observed Compton-$y$ field as
\begin{equation}
y_W (\bm{\theta}) = W(\bm{\theta}) y(\bm{\theta}),
\end{equation}
where $W(\bm{\theta}) = 1$ for observed regions and $W(\bm{\theta}) = 0$ for masked regions.
Then, the Fourier transform of the field under flat-sky approximation is given by
\begin{equation}
\tilde{y}_W (\bm{\ell}) = \int \frac{\mathrm{d}^2 \ell'}{(2 \pi)^2} \tilde{W} (\bm{\ell} - \bm{\ell}') \tilde{y} (\bm{\ell}') ,
\end{equation}
where the tilde symbol denotes the Fourier transform of the quantity.
The estimator of the power spectrum for $i^\mathrm{th}$ multipole bin is given as
\begin{equation}
\hat{C}(\ell_i) \equiv \frac{1}{\Omega_W} \int_{\ell \in \ell_i}
\frac{\mathrm{d}^2 \ell}{\Omega_{\ell_i}} \tilde{y}_W (\bm{\ell}) \tilde{y}_W (-\bm{\ell}) ,
\end{equation}
where $\Omega_W$ is the effective survey area
$\Omega_W = \int \mathrm{d}^2 \theta \, W(\bm{\theta})$,
and $\Omega_{\ell_i}$ is the number of modes
$\Omega_{\ell_i} = \int_{\ell \in \ell_i} \mathrm{d}^2 \ell$.
When the bin width is sufficiently smaller than the multipole,
the number of modes can be approximated as
\begin{equation}
\Omega_{\ell_i} \approx 2 \pi \ell_i^2 \Delta \ln \ell ,
\end{equation}
where we consider the logarithmically
evenly-spaced bins for multipoles
and $\Delta \ln \ell$ is the bin width in logarithmic space.
Then, the ensemble average of the estimator becomes
the sum of the power spectrum $C(\ell)$ and noise power spectrum $N(\ell)$:
\begin{equation}
\langle \hat{C} (\ell_i) \rangle = C (\ell_i) + N (\ell_i) .
\end{equation}
Hereafter, the parenthesis denotes ensemble average.

\subsection{Covariance Matrix of the tSZ Power Spectrum}
Next, we derive the expression for the covariance matrix $\mathcal{C}$ as
\begin{align}
\mathcal{C}_{ij} &\equiv \langle \hat{C} (\ell_i) \hat{C} (\ell_j) \rangle -
\langle \hat{C} (\ell_i) \rangle \langle \hat{C} (\ell_j) \rangle \nonumber \\
&= \frac{1}{\Omega_W} \left[ \frac{(2 \pi)^2}{\Omega_{\ell_i}} 2 [C (\ell_i) + N (\ell_i)]^2 \delta_{ij}
+ \bar{T}_W (\ell_i, \ell_j) \right] ,
\end{align}
where $\delta_{ij}$ is the Kronecker delta.
The first term is referred to as
the Gaussian term (labeled as $\mathrm{G}$):
\begin{equation}
\mathcal{C}^\mathrm{G}_{ij} = \frac{1}{\Omega_W} \frac{(2 \pi)^2}{\Omega_{\ell_i}}
2 [C (\ell_i) + N (\ell_i)]^2 \delta_{ij} .
\end{equation}
The remaining term is the non-Gaussian term, which does not depend on the binning of the multipoles
in contrast to the Gaussian term.
The trispectrum convoluted with the window function $\bar{T}_W$ is given as
\begin{widetext}
\begin{align}
\bar{T}_W (\ell_i, \ell_j) = \frac{1}{\Omega_W} \int_{l \in l_i}
\frac{\mathrm{d}^2 \ell}{\Omega_{\ell_i}} \int_{l \in l_j} \frac{\mathrm{d}^2 \ell'}{\Omega_{\ell_j}}
\int \left[ \prod_{a=1}^4 \frac{\mathrm{d}^2 q_a}{(2 \pi)^2} \tilde{W} (\bm{q}_a) \right]
(2\pi)^2 \delta_\mathrm{D} (\bm{q}_{1234})
T(\bm{\ell} + \bm{q}_1, -\bm{\ell}+\bm{q}_2, \bm{\ell}'+\bm{q}_3, -\bm{\ell}'+\bm{q}_4) ,
\end{align}
\end{widetext}
where $\delta_\mathrm{D}$ is the Dirac delta function and
$\bm{q}_{1234} \equiv \bm{q}_1 + \bm{q}_2 + \bm{q}_3 + \bm{q}_4$.

Then, we derive the expressions of the covariance matrix
based on halo model \cite{Cooray2001}.
First, we introduce the following notation:
\begin{equation}
I_\mu^\beta (\ell_1, \ldots , \ell_\mu ;\chi) \equiv
\int \mathrm{d}M \frac{\mathrm{d}n_\mathrm{h}}{\mathrm{d}M} b_\beta \tilde{y} (\ell_1) \cdots \tilde{y} (\ell_\mu) ,
\end{equation}
where $b_0 = 1$ and $b_1 = b_\mathrm{h} (M, z)$.
The halo model expression of the tSZ power spectrum becomes
\begin{align}
C (\ell) = \int \mathrm{d} \chi \frac{\mathrm{d}^2 V}{\mathrm{d}\chi \mathrm{d}\Omega}
& \biggl[ I_2^0 (\ell, \ell ; \chi)  + [I_1^1 (\ell ; \chi)]^2
\nonumber \\
\label{eq:hm_cl}
& \times P_\mathrm{L} \left( k = \frac{\ell+1/2}{\chi} , \chi \right) \biggl].
\end{align}
For projected fields such as Compton-$y$, we can compute
the non-Gaussian terms for covariance as follows \cite{Takada2007,Takada2009,Sato2009}:
\begin{widetext}
\begin{align}
\mathcal{C}_{ij}^\mathrm{NG} &= \mathcal{C}_{ij}^\mathrm{cNG} + \mathcal{C}_{ij}^\mathrm{SSC}, \\
\label{eq:hm_cNG}
\mathcal{C}_{ij}^\mathrm{cNG} &= \frac{1}{\Omega_W} \int_{l \in l_i}
\frac{\mathrm{d}^2 \ell}{\Omega_{\ell_i}} \int_{l \in l_j} \frac{\mathrm{d}^2 \ell'}{\Omega_{\ell_j}}
T(\bm{\ell}, -\bm{\ell}, \bm{\ell}', -\bm{\ell}'), \\
\mathcal{C}_{ij}^\mathrm{SSC} &=
\mathcal{C}_{ij}^\mathrm{HSV} +
\mathcal{C}_{ij}^\mathrm{HSV\text{-}BC} + \mathcal{C}_{ij}^\mathrm{BC} , \\
\mathcal{C}_{ij}^\mathrm{HSV} &= \int \mathrm{d}\chi \frac{\mathrm{d}^2 V}{\mathrm{d}\chi \mathrm{d}\Omega}
I^1_2 (\ell_i, \ell_i ;\chi) I^1_2 (\ell_j, \ell_j ;\chi)
\frac{1}{\Omega_W^2} \int \frac{\mathrm{d}^2 \ell}{(2 \pi)^2}
|\tilde{W} (\bm{\ell})|^2 P_\mathrm{L} (k; \chi) \nonumber \\
\label{eq:hm_HSV}
&=  \int \mathrm{d}\chi \frac{\mathrm{d}^2 V}{\mathrm{d}\chi \mathrm{d}\Omega}
I^1_2 (\ell_i, \ell_i ;\chi) I^1_2 (\ell_j, \ell_j ;\chi) [\sigma_W^L (\chi)]^2 , \\
\label{eq:hm_BC}
\mathcal{C}_{ij}^\mathrm{BC} &= \int \mathrm{d}\chi \frac{\mathrm{d}^2 V}{\mathrm{d}\chi \mathrm{d}\Omega}
\left( \frac{68}{21} \right)^2 [I_1^1 (\ell_i; \chi) I_1^1 (\ell_j; \chi)]^2
P_\mathrm{L} (k_i; \chi) P_\mathrm{L} (k_j; \chi) [\sigma_W^L (\chi)]^2 , \\
\label{eq:hm_HSV-BC}
\mathcal{C}_{ij}^\mathrm{HSV\text{-}BC} &= \int \mathrm{d}\chi \frac{\mathrm{d}^2 V}{\mathrm{d}\chi \mathrm{d}\Omega}
\frac{68}{21}
\left\{ [I_1^1 (\ell_i ;\chi)]^2 I_2^1 (\ell_j, \ell_j; \chi) P_\mathrm{L} (k_i ;\chi) +
[I_1^1 (\ell_j ;\chi)]^2 I_2^1 (\ell_i, \ell_i; \chi) P_\mathrm{L} (k_j ;\chi) \right\}
[\sigma_W^L (\chi)]^2 ,
\end{align}
\end{widetext}
where we define the variance of mass density fluctuations of super-survey modes as
\begin{equation}
[\sigma_W^L (\chi)]^2 \equiv \frac{1}{\Omega_W^2} \int \frac{\mathrm{d}^2 \ell}{(2 \pi)^2} |\tilde{W} (\bm{\ell})|^2 P_\mathrm{L} (k; \chi) ,
\end{equation}
and $k \equiv (\ell+1/2)/\chi$.
The term which is sourced from the trispectrum with parallelogram configuration is referred to as
the connected non-Gaussian term (labeled as $\mathrm{cNG}$).
The latter three terms are referred to as
the halo sample variance \citep[][HSV]{Sato2009}
the beat coupling \citep[][BC]{Hamilton2006,Takada2009}
and their cross-correlation (HSV-BC) terms, respectively.
The sum of the three terms is referred to as super sample variance  \citep[][SSC]{Takada2013}.
Since the trispectrum is dominated by the one-halo term at all scales \cite{Cooray2001},
we ignore the two-, three-, and four-halo terms of the trispectrum for simplicity:
\begin{align}
T(\bm{\ell}, -\bm{\ell}, \bm{\ell}', -\bm{\ell}') \approx & T^\mathrm{1h} (\ell, \ell')
\nonumber \\
= & \int \mathrm{d}\chi \frac{\mathrm{d}^2 V}{\mathrm{d}\chi \mathrm{d}\Omega}
\int \mathrm{d}M \frac{\mathrm{d}n_\mathrm{h}}{\mathrm{d}M}
\nonumber \\
& \times |\tilde{y} (\ell ;M, \chi)|^2 |\tilde{y} (\ell'; M, \chi)|^2 .
\end{align}

\subsection{Implementation of Cluster Masking in Halo Model}
\label{sec:masking_hm}
Here, we describe how to incorporate the cluster masking in the halo model expressions
of tSZ power spectrum and covariance matrix.
The selection function $S(M, \chi)$ (Eq.~\ref{eq:selection}) is inserted in the mass integration
in $I_\mu^\beta$ and $T^\mathrm{1h}$:
\begin{align}
I_\mu^\beta (\ell_1, \ldots , \ell_\mu ;\chi) & \to
\hat{I}_\mu^\beta (\ell_1, \ldots , \ell_\mu ;\chi)
\nonumber \\
 & \equiv \int \mathrm{d}M \frac{\mathrm{d}n_\mathrm{h}}{\mathrm{d}M} S(M, \chi)
b_\beta \tilde{y} (\ell_1) \cdots \tilde{y} (\ell_\mu),
\end{align}
and
\begin{align}
T^\mathrm{1h} (\ell, \ell') \to &
\hat{T}^\mathrm{1h} (\ell, \ell')
\nonumber \\
\equiv & \int \mathrm{d}\chi \frac{\mathrm{d}^2 V}{\mathrm{d}\chi \mathrm{d}\Omega}
\int \mathrm{d}M \frac{\mathrm{d}n_\mathrm{h}}{\mathrm{d}M} S(M, \chi)
\nonumber \\
& \times |\tilde{y} (\ell ;M, \chi)|^2 |\tilde{y} (\ell'; M, \chi)|^2 .
\end{align}
Then, in order to apply cluster masking,
$I_\mu^\beta$ and $T^\mathrm{1h}$ in halo model expressions
(Eqs.~\ref{eq:hm_cl}, \ref{eq:hm_cNG}, \ref{eq:hm_HSV}, \ref{eq:hm_BC}, and \ref{eq:hm_HSV-BC})
are replaced with $\hat{I}_\mu^\beta$ and $\hat{T}^\mathrm{1h}$, respectively.

\subsection{Weight Function}
\label{sec:weight}
In order to investigate which halos contribute the signal and covariance,
we compute the weight function with respect to mass and redshift.
For the power spectrum, the weight function is given as
\begin{align}
\frac{\mathrm{d}^2}{\mathrm{d}z \mathrm{d}M} C(\ell) = & \frac{\mathrm{d}^2 V}{\mathrm{d}z \mathrm{d}\Omega}
\left[ \frac{\mathrm{d}n_\mathrm{h}}{\mathrm{d}M} |\tilde{y} (\ell)|^2
+ 2 \frac{\mathrm{d}n_\mathrm{h}}{\mathrm{d}M} b_\mathrm{h} \tilde{y} (\ell) \right.
\nonumber \\
& \left. \times P_\mathrm{L} \left( k = \frac{\ell+1/2}{\chi} , z \right)
\int \mathrm{d}M \frac{\mathrm{d}n_\mathrm{h}}{\mathrm{d}M} b_\mathrm{h} \tilde{y} (\ell) \right] .
\end{align}
Similarly, the weight function of the trispectrum is given as
\begin{align}
\frac{\mathrm{d}^2}{\mathrm{d}z \mathrm{d}M} T(\bm{\ell}, -\bm{\ell}, \bm{\ell}', -\bm{\ell}') & \approx
\frac{\mathrm{d}^2}{\mathrm{d}z \mathrm{d}M} T^\mathrm{1h} (\ell, \ell') \nonumber \\
& = \frac{\mathrm{d}^2 V}{\mathrm{d}z \mathrm{d}\Omega}
\frac{\mathrm{d}n_\mathrm{h}}{\mathrm{d}M} |\tilde{y} (\ell)|^2 |\tilde{y} (\ell')|^2 .
\end{align}
Note that these expressions are weight functions with respect to the redshift $z$,
instead of the comoving distance $\chi$,
and the comoving volume with respect to redshift
is $\mathrm{d}^2 V/\mathrm{d}z \mathrm{d}\Omega = \chi^2 (z) H (z)/c$.

\begin{figure*}[htbp]
  \centering
  \includegraphics[width=\textwidth,clip]{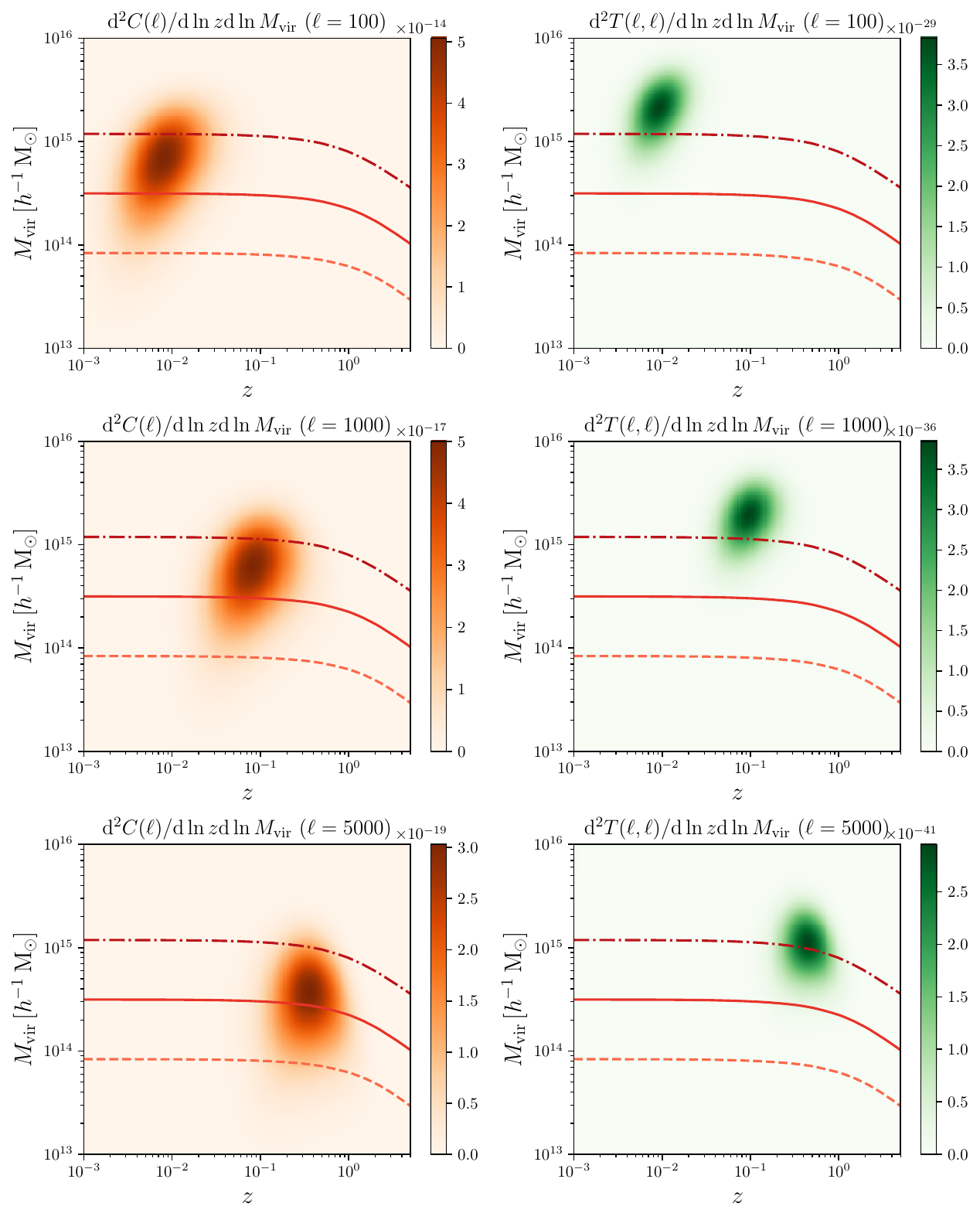}
  \caption{The differential contributions of clusters in infinitesimal intervals of redshift and halo mass around given redshift ($x$-axis) and mass ($y$-axis) to
  the power spectrum and the
  diagonal term of trispectrum for $\ell = 100$ (upper panels),
  $1000$ (medium panels), $5000$ (lower panels), respectively.
  The dashed, solid, and dot-dashed lines correspond to $\log (Y_{500}/\mathrm{Mpc}^2)
  = -6, -5, -4$, respectively.}
  \label{fig:dCdT}
\end{figure*}

Figure~\ref{fig:dCdT} shows the weight functions for diagonal components, i.e., $\ell = \ell'$
of power spectrum and trispectrum for three representative multipoles $\ell = 100$, $1000$, $5000$.
Obviously, the trispectrum is more sensitive to massive clusters than power spectrum at all scales.
The integrand of mass integration of trispectrum
contains $\tilde{y}^4$ in contrast to $\tilde{y}^2$ at highest for power spectrum,
and thus, the contribution from massive clusters becomes prominent in trispectrum.
Similarly, we can expect that
the super-sample covariance involves
the mass integration of $\tilde{y}^2$ at highest
and is less sensitive to massive clusters compared with the cNG term.
We also show the three critical lines of $\log (Y_{500}/\mathrm{Mpc}^2)
= -6$ (dashed), $-5$ (solid), $-4$ (dot-dashed).
In order to enhance the significance of the detection,
we need to keep the power spectrum but remove the trispectrum contribution
as much as possible, which is the dominant source of the covariance.
We adopt the fiducial threshold as
$\log (Y_\mathrm{thres}/\mathrm{Mpc}^2) = -5$ when masking massive clusters.
Apparently, the threshold $\log (Y_\mathrm{thres}/\mathrm{Mpc}^2) = -4$
keeps the large fraction of the signal but
our fiducial threshold gives higher significance.
Since the contribution of the cNG term is quite larger than other terms,
rather than maintaining the signal,
completely excluding the trispectrum is a more effective strategy.

\section{Experimental Conditions}
\label{sec:experiment}
\begin{figure}[htbp]
  \centering
  \includegraphics[width=\columnwidth,clip]{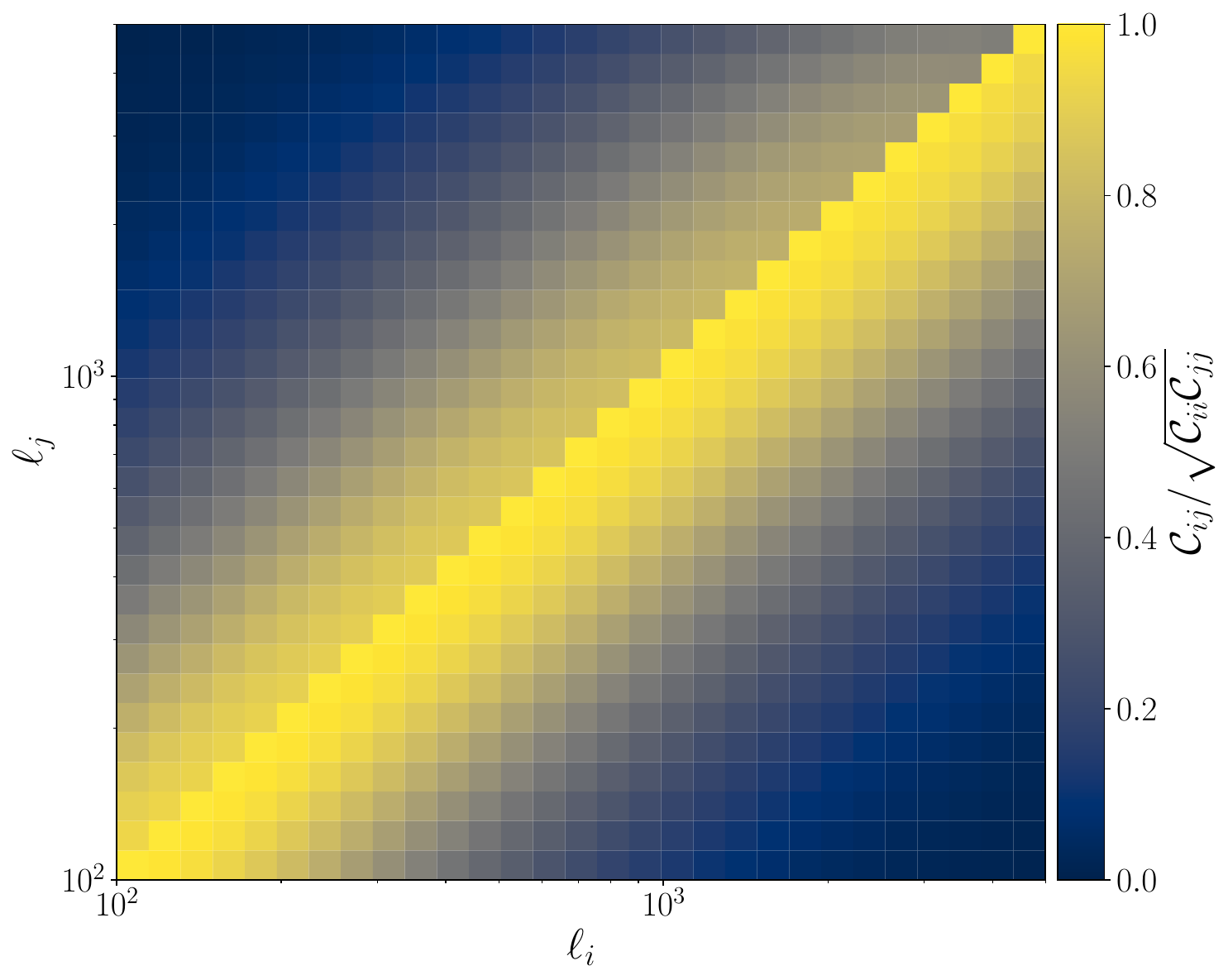}
  \caption{The total covariance matrix of the tSZ power spectrum.
  The upper left (lower right) part corresponds to the covariance matrix
  with (without) cluster masking.}
  \label{fig:cov}
\end{figure}

\begin{figure*}[htbp]
  \centering
  \includegraphics[width=\textwidth,clip]{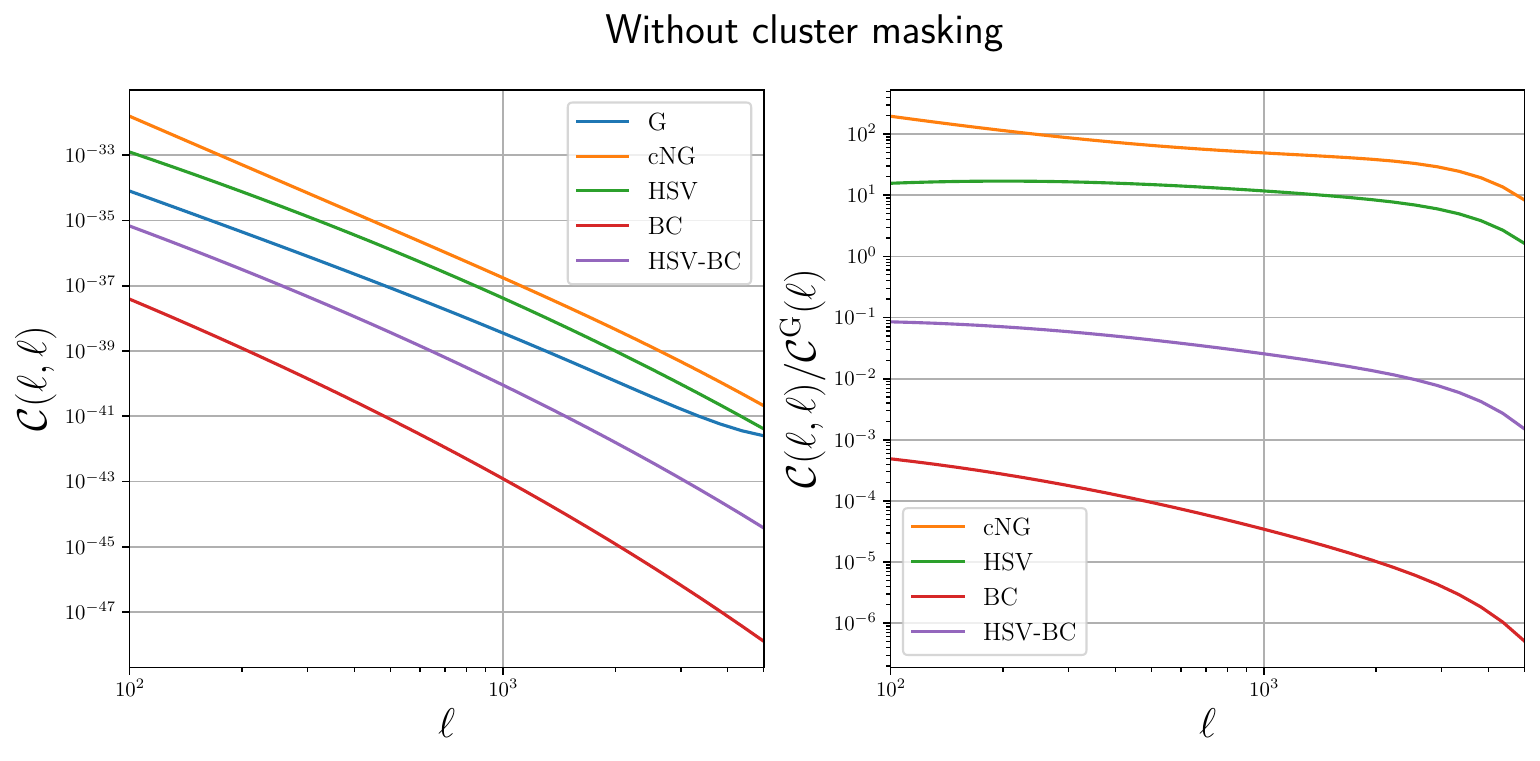}
  \includegraphics[width=\textwidth,clip]{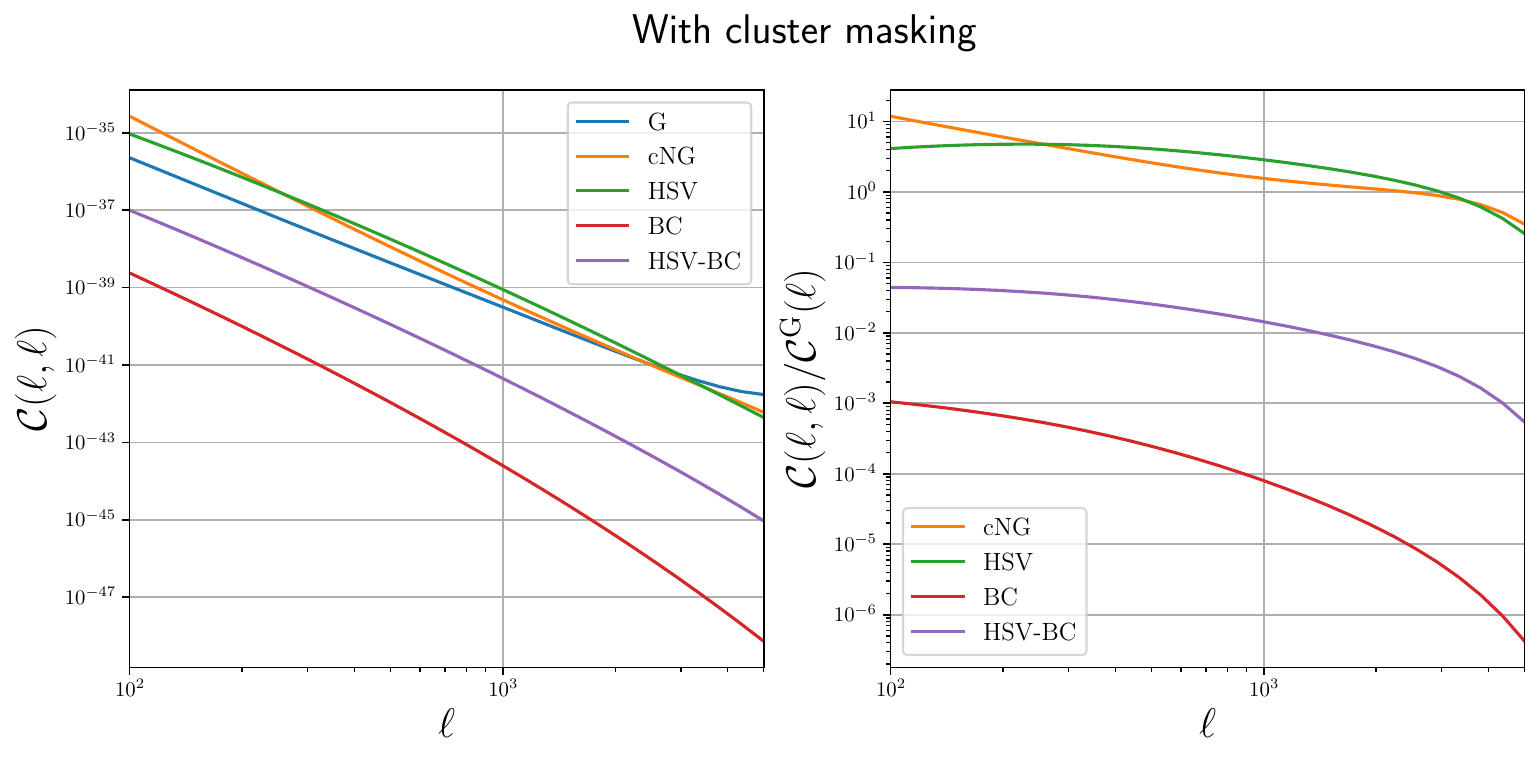}
  \caption{The diagonal terms of the covariance matrix of the tSZ power spectrum (left panels)
  and each term divided by the Gaussian term (right panels)
  without (upper panels) and with (lower panels) cluster masking.
  The HSV, BC, and HSV-BC terms correspond to the SSC.}
  \label{fig:cov_diag}
\end{figure*}

In this Section, we define the experimental conditions for statistical analysis.
We consider a practical case which is similar to
Advanced ACT measurement \cite{Madhavacheril2020}.
We assume the sky coverage is $\Omega_W = 2100 \, \mathrm{deg}^2$ and
the survey window is circular symmetric for simplicity.
The power spectrum of the mask is given as
\begin{equation}
|\tilde{W} (\bm{\ell})|^2 = \Omega_W^2 \left[ 2 \frac{J_1 (\ell \Theta_W)}{\ell \Theta_W} \right]^2 ,
\end{equation}
where $J_1 (x)$ is the first-order Bessel function and
we employ $\Theta_W \equiv \sqrt{\Omega_W/\pi}$.
In general, the survey window function has irregular shape
or the survey regions are divided into multiple separate patches.
The relative strength of SSC depends on the geometry of the survey region or
the degree of discontinuity of survey regions \cite{Takahashi2014}.
The compact survey geometry such as a circular geometry considered here has the largest
SSC contribution. Hence the following estimate can be considered as
the worst case scenario of the impact of SSC contribution.
The binning is logarithmically evenly
spaced with the minimum $\ell_\mathrm{min} = 100$,
the maximum $\ell_\mathrm{max} = 5000$, and the number of bins $n_\ell = 30$.
For the noise power spectrum, we assume the instrumental noise is Gaussian
and its variance is $\sigma_\mathrm{inst} = 7 \, \mu \mathrm{K} \, \mathrm{arcmin}$
with the single band at $\nu_0 = 149\,\mathrm{GHz}$.
Then, the noise power spectrum is given as
\begin{equation}
N (\ell) = \left[
\frac{\sigma_\mathrm{inst}}{g(\nu_0) T_\mathrm{CMB}} \right]^2
e^{\ell^2 \theta^2_\mathrm{FWHM} / (8 \ln 2)} ,
\end{equation}
where $\theta_\mathrm{FWHM} = 1.4 \, \mathrm{arcmin}$ is
the full width at half maximum of the beam size \cite{Madhavacheril2020}.
In addition to the instrumental noise, radio and infrared point sources,
cosmic infrared background, and primary CMB leak to Compton-$y$ estimates
due to incomplete separation and can be a source of noise.
However, the instrumental noise dominates at small scales \cite{Hill2013,Bolliet2018},
and we ignore the contributions from other sources in the subsequent analyses for simplicity.
The total covariance matrix is shown in Figure~\ref{fig:cov}.
The diagonal term of the covariance matrix for each component
is shown in Figure~\ref{fig:cov_diag}.
In the case of no cluster masking,
the cNG term dominates at all scales.
On the other hand, cluster masking removes clusters which contribute to the cNG term,
and Gaussian and SSC terms become important when clusters are masked.

\section{Results}
\label{sec:results}

\subsection{Statistical Significance}
\label{sec:SN}
\begin{figure}[htbp]
  \centering
  \includegraphics[width=\columnwidth,clip]{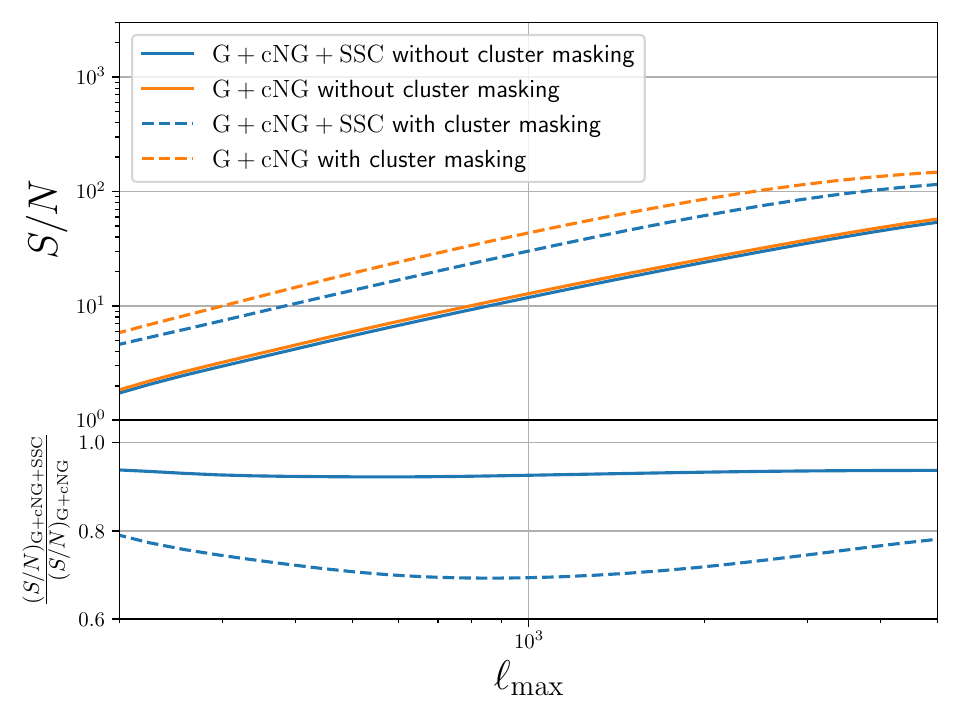}
  \caption{The SNR as a function of maximum multipole $\ell_\mathrm{max}$
  with the total covariance including SSC or
  the covariance including only the Gaussian
  and trispectrum (cNG) contributions (i.e. without SSC), as denoted in the legend.
  The lowest multipole is fixed as $\ell_\mathrm{min} = 100$
  and the highest multipole is varied in the range of $200 \leq \ell_\mathrm{max} \leq 5000$.
  The bin width is also fixed.
  The lower panel shows the ratio of $S/N$ values with and without the SSC contribution,
  for the two cases where we remove massive clusters with $Y_{500} > 10^{-5}\, \mathrm{Mpc}^{-2}$
  or not (with and without ``cluster masking'').}
  \label{fig:SN}
\end{figure}

In this section, we discuss the statistical significance
of the detection of the tSZ power spectrum and the effects of
cluster masking and SSC on it.
The signal-to-noise ratio (SNR) $S/N$ is computed as
\begin{equation}
\frac{S}{N} =
\left[ \sum_{\ell_\mathrm{min} \leq \ell_i, \ell_j \leq \ell_\mathrm{max}}
C (\ell_i) \mathcal{C}^{-1}_{ij} C (\ell_j) \right]^{\frac{1}{2}}.
\end{equation}
Figure~\ref{fig:SN} shows the SNR
as a function of the maximum multipole $\ell_\mathrm{max}$.
The SSC contributes to only $5\%$ of the SNR without cluster masking
because the cNG term dominates.
However, by masking massive clusters, the overall SNR is enhanced
due to the suppression of the cNG term but
the relative contribution from SSC also rises up to $30\%$ because
the SSC is less sensitive to the cluster masking.

Next, we discuss how SSC affects
the chi-square statistic that an observer could obtain for a given realization of data.
The chi-square statistic is defined as
\begin{equation}
\chi^2 \equiv (\bm{d}-\bar{\bm{d}})^T \mathcal{C}^{-1} (\bm{d}-\bar{\bm{d}}) ,
\end{equation}
where $\bm{d}$ is the data vector defined as
\begin{equation}
\bm{d} = (C(\ell_1), \ldots, C(\ell_{n_\ell})) ,
\end{equation}
and $\bar{\bm{d}} \equiv \langle \bm{d} \rangle$.
The expectation value of the $\chi^2$ value is estimated as
\begin{align}
\langle \chi^2 \rangle & = \langle (\bm{d}-\bar{\bm{d}})^\mathrm{T} \mathcal{C}^{-1}
(\bm{d}-\bar{\bm{d}}) \rangle \nonumber \\
& = \mathrm{Tr} ( \langle (\bm{d}-\bar{\bm{d}}) (\bm{d}-\bar{\bm{d}})^\mathrm{T} \rangle
\mathcal{C}^{-1} ) \nonumber \\
& = \mathrm{Tr} (\mathcal{C} \mathcal{C}^{-1}) = n_\ell .
\end{align}
However, if the wrong covariance is employed in the analysis,
the expected chi-square deviates from the number of bins.
Let us consider the case where
the true covariance is composed of Gaussian, connected non-Gaussian,
and super-sample covariance, and the wrong covariance is that
without the super-sample covariance:
\begin{align}
  \mathcal{C}_\text{true} \equiv & \mathcal{C}^\mathrm{G} + \mathcal{C}^\mathrm{cNG} +
  \mathcal{C}^\mathrm{SSC}, \\
  \mathcal{C}_\text{wrong} \equiv & \mathcal{C}^\mathrm{G} + \mathcal{C}^\mathrm{cNG} .
\end{align}
Then, the expectation value of chi-square with the wrong covariance is given as
\begin{align}
\langle \chi^2_\text{wrong} \rangle & =
\langle (\bm{d}-\bar{\bm{d}})^\mathrm{T} (\mathcal{C}_\text{wrong})^{-1}
(\bm{d}-\bar{\bm{d}}) \rangle \nonumber \\
& = \mathrm{Tr} \left( \mathcal{C}_\text{true}
(\mathcal{C}_\text{wrong})^{-1}
\right) > n_\ell .
\end{align}
The chi-square with the wrong covariance no longer follows
the chi-squared distribution but if the wrong covariance is close to
the true covariance, we can approximate
the probability distribution function (PDF)
of $\chi^2_\text{wrong}$ as the chi-squared distribution with scaling to match
the expectation value:
\begin{equation}
P(\chi^2_\text{wrong}) \mathrm{d}\chi^2_\text{wrong} \approx
P_{\chi^2} \left( \frac{n_\ell}{\alpha} \chi^2_\text{wrong}; n_\ell \right)
\frac{n_\ell}{\alpha} \mathrm{d}\chi^2_\text{wrong},
\end{equation}
where $P_{\chi^2} (\chi^2; n)$ is PDF of chi-squared distribution
with the degree of freedom $n$, and
$\alpha \equiv \mathrm{Tr} \left( \mathcal{C}_\text{true}
(\mathcal{C}_\text{wrong})^{-1} \right)$.

\begin{figure}[htbp]
  \centering
  \includegraphics[width=\columnwidth,clip]{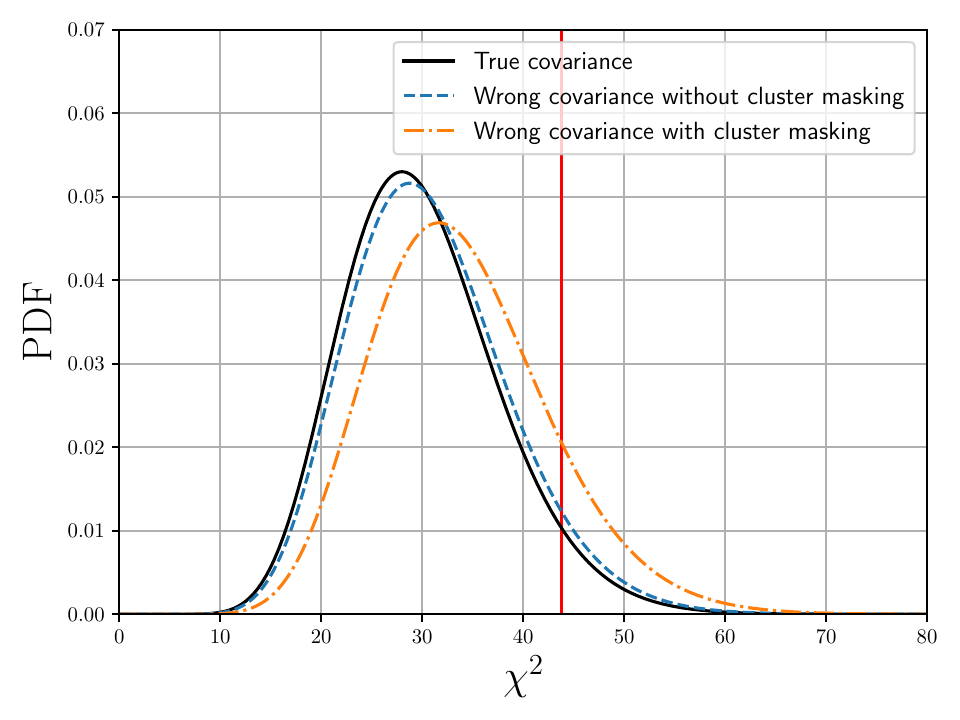}
  \caption{The PDFs of chi-square
  with true and wrong covariances
  and with and without cluster masking.
  The black solid line shows the chi-square distribution
  with degree of freedom $n = 30$ and for the other lines,
  the argument is scaled to match the expectation values.
  The blue dashed line shows the PDF
  with the wrong covariance, i.e., without the SSC term,
  and furthermore, the orange dot-dashed line shows the one
  when massive clusters are masked.
  The red vertical line corresponds to $\chi^2 = 43.8$
  where the $p$-value is $0.05$
  with the true covariance.}
  \label{fig:chi_square}
\end{figure}

Figure~\ref{fig:chi_square} shows the PDFs with the true covariance
and the wrong ones, i.e., without SSC, for the cases
with and without cluster masking.
The PDFs with wrong covariances are skewed rightward compared with the true one
because the wrong covariance underestimates the true covariance
or more exactly the amplitude of statistical scatters,
and it apparently gives a higher significance.
In order to see how the significance is overestimated with the wrong covariance,
we address how the $p$-value changes.
First, we compute the $\chi^2_{p = 0.05}$,
which gives the $p$-value as $0.05$ with the true covariance.
When the true covariance is adopted, the resultant chi-square follows
the chi-squared distribution and from the cumulative distribution function,
$\chi^2_{p = 0.05} = 43.8$ is calculated.
Then, we compute the $p$-values with wrong covariances, i.e.,
\begin{equation}
p = \int_{\chi^2_{p = 0.05}}^\infty \!\! P(\chi^2_\text{wrong})
\, \mathrm{d} \chi^2_\text{wrong} .
\end{equation}
In Table~\ref{tab:p-values}, we show the $p$-values
for the two cases with and without cluster masking.
When cluster masking is not applied,
the effect of the super-sample covariance is subdominant
and the $p$-value increases only by $1.1\%$.
However, when massive clusters are masked,
the SSC becomes relatively important and
the $p$-value is $12.0\%$.
This result implies that it is $7.0\%$ more likely to derive
optimistic significance of the detection.
Or for a case of parameter inference,
even a true model gives an apparent bad fitting to the data,
because there might be a higher chance to have
a relatively large chi-square value for the best-fit model
due to the underestimation in the covariance amplitude.

\begin{table}
  \begin{ruledtabular}
    \begin{tabular}{lcc}
      Covariance & $\langle \chi^2 \rangle$ & $p$-value \\
      \hline
      True covariance & $30$ & $0.05$ \\
      Wrong covariance without cluster masking & $30.80$ & $0.061$ \\
      Wrong covariance with cluster masking & $33.91$ & $0.120$ \\
    \end{tabular}
  \end{ruledtabular}
  \caption{The $p$-values with wrong covariances
  with and without cluster masking.
  The lower limit of chi-square is determined as the $p$-value with the true covariance is $0.05$.}
  \label{tab:p-values}
\end{table}

\subsection{Fisher Analysis}
In this Section, we quantify the effects of SSC on the parameter constraints with
tSZ power spectrum based on Fisher forecast \citep{Tegmark1997a,Tegmark1997b}.
The tSZ power spectrum is sensitive to the matter fluctuation and
has potential to place tight constraints on the amplitude $A_\mathrm{s}$.
Furthermore, the hydrostatic mass bias parameter can also be constrained,
which information is not accessible solely from CMB temperature and polarization analysis.

\begin{figure*}[htbp]
  \centering
  \includegraphics[width=\textwidth,clip]{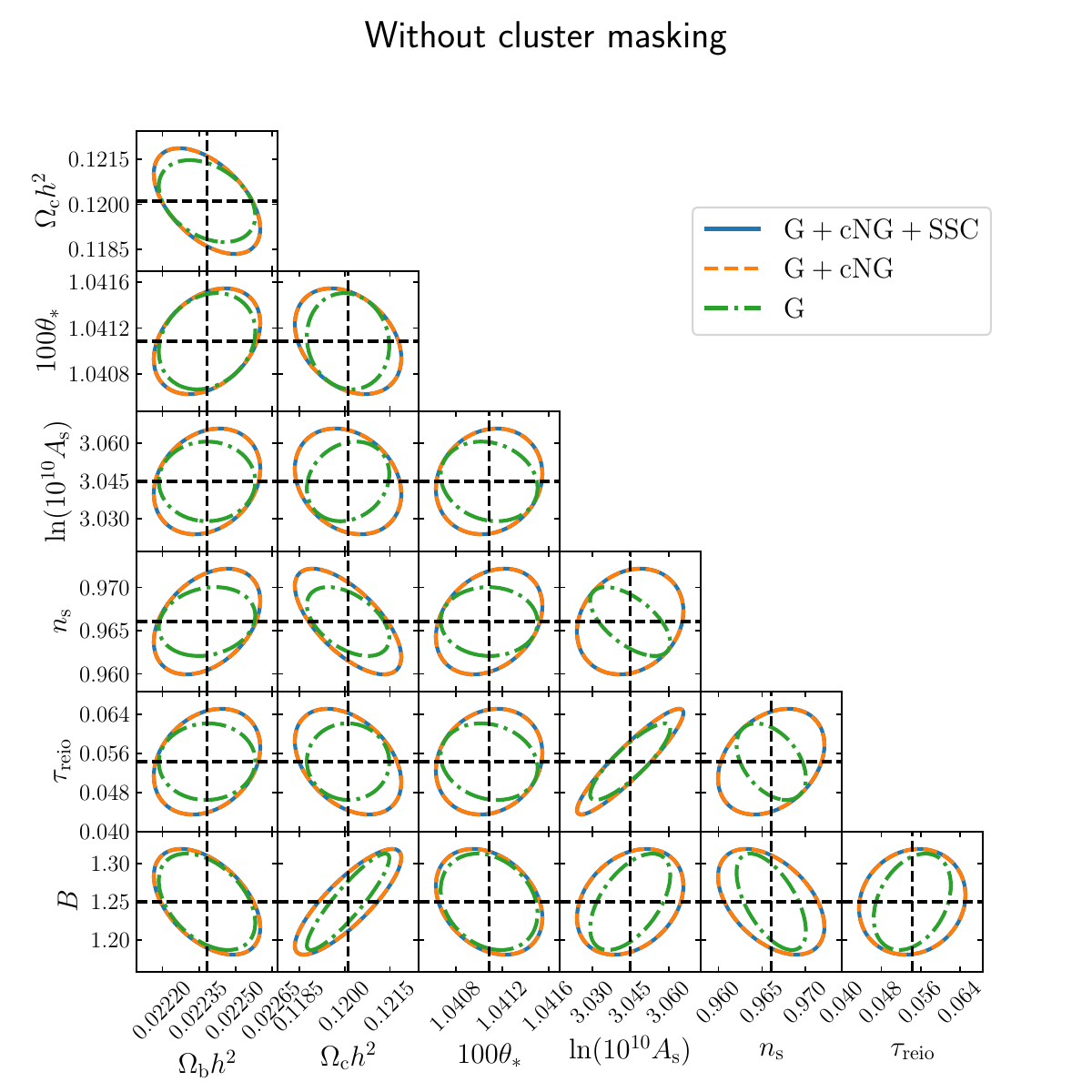}
  \caption{The expected $1\text{-}\sigma$ constraints
  on parameters including marginalization over other parameters,
  obtained based on the Fisher matrix (see text for details).
  The cluster masking is not applied.
  The blue solid, orange dashed, and green dot-dashed lines
  correspond to the results with total, Gaussian and connected non-Gaussian,
  Gaussian only covariances, respectively.}
  \label{fig:region_nocut}
\end{figure*}

\begin{figure*}[htbp]
  \centering
  \includegraphics[width=\textwidth,clip]{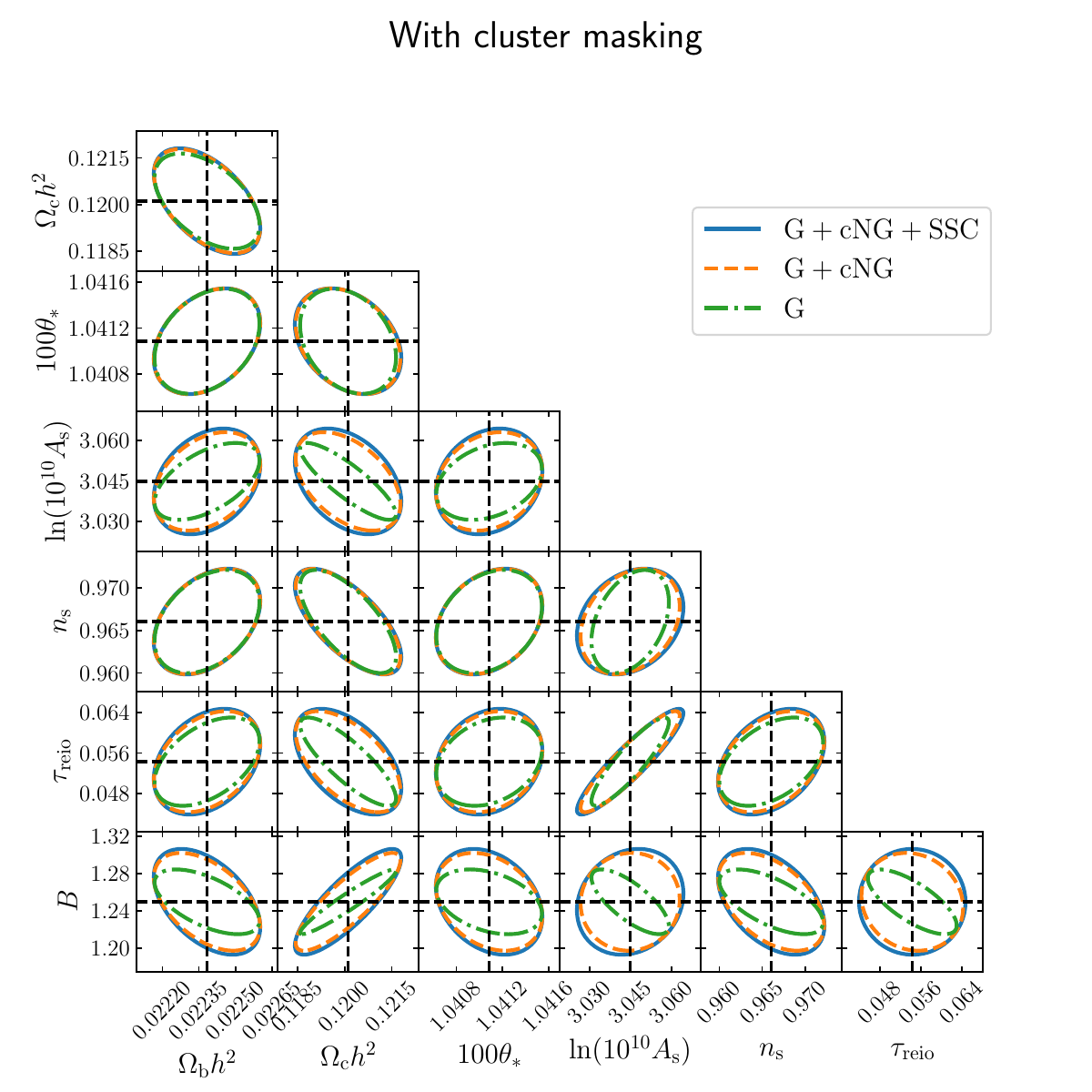}
  \caption{Similar to the previous figure, but the expected constraints when
  the cluster masking is applied.
  The blue solid, orange dashed, and green dot-dashed lines
  correspond to the results with total, Gaussian and connected non-Gaussian,
  Gaussian only covariances, respectively.}
  \label{fig:region_Ycut}
\end{figure*}

When the likelihood is assumed to be multi-variate Gaussian,
the Fisher matrix is given as
\begin{equation}
F^{\mathrm{tSZ}}_{ij} = \frac{\partial \bm{d}^\mathrm{T}}{\partial \theta_i} \mathcal{C}^{-1}
\frac{\partial \bm{d}}{\partial \theta_j}
+\frac{1}{2} \mathrm{Tr} \left(\mathcal{C}^{-1}
\frac{\partial \mathcal{C}}{\partial \theta_i}
\mathcal{C}^{-1}
\frac{\partial \mathcal{C}}{\partial \theta_j}
\right) .
\end{equation}
We consider the parameter space
$\bm{\theta} = (\Omega_\mathrm{b} h^2, \Omega_\mathrm{c} h^2, 100 \theta_*,
\ln (10^{10} A_\mathrm{s}), n_\mathrm{s}, \tau_\mathrm{reio}, B)$,
where $B = (1-b_\mathrm{HSE})^{-1}$.
In particular, the hydrostatic bias parameter $B$ characterizes
the non-thermal pressure support
of galaxy clusters and groups,
and it is not well constrained compared with cosmological parameters.
Since only with the tSZ power spectrum, severe degeneracy between parameters occurs,
we add information from
\textit{Planck} 2018 TT,TE,EE+lowE+lensing result \cite{Planck2018VI}.
Thus, the resultant Fisher matrix $F$ is given by
\begin{equation}
F = F^\mathrm{tSZ} + F^\mathrm{CMB} ,
\end{equation}
where the Fisher matrix for CMB anisotropies are computed from
obtained chains in \textit{Planck} analysis.
In Figures~\ref{fig:region_nocut} and \ref{fig:region_Ycut},
forecasts of the constraints from tSZ power spectrum and
CMB anisotropies with and without cluster masking are shown.
Table~\ref{tab:constraints} shows the forecast of $1\text{-}\sigma$ errors.
When adding the information of the tSZ power spectrum,
we can obtain tighter constraints on parameters, especially
the amplitude of primordial curvature perturbations:
$\ln (10^{10} A_\mathrm{s})$.
Note that the hydrostatic bias parameter $B$ can also be tightly constrained.
When clusters are not masked, the effect of SSC is subdominant since the cNG term is much larger.
On the other hand, with cluster masking, the SSC comes into effect
because the SSC is not sensitive to cluster masking compared with the cNG term.
As the SNR improves by cluster masking, the constraints become tighter.

\begin{table*}
  \begin{ruledtabular}
    \begin{tabular}{lccccccc}
      & $\Omega_\mathrm{b} h^2$ & $\Omega_\mathrm{c} h^2$ & $100 \theta_*$ &
      $\ln (10^{10} A_\mathrm{s})$ & $n_\mathrm{s}$ & $\tau_\mathrm{reio}$ & $B$ \\
      \hline
      CMB & $0.000146$ & $0.00118$ & $0.000307$ & $0.0141$ & $0.00415$ & $0.00730$ & --- \\
      \hline
      \multicolumn{8}{c}{Without cluster masking} \\
      CMB+tSZ ($\mathrm{G}$) & $0.000132$ & $0.00090$ & $0.000278$ & $0.0105$ & $0.00264$ & $0.00520$ & $0.0416$ \\
      CMB+tSZ ($\mathrm{G}+\mathrm{cNG}$) & $0.000145$ & $0.00117$ & $0.000305$ & $0.0139$ & $0.00404$ & $0.00716$ & $0.0455$ \\
      CMB+tSZ ($\mathrm{G}+\mathrm{cNG}+\mathrm{SSC}$) & $0.000145$ & $0.00117$ & $0.000305$ & $0.0139$ & $0.00405$ & $0.00717$ & $0.0456$ \\
      \hline
      \multicolumn{8}{c}{With cluster masking} \\
      CMB+tSZ ($\mathrm{G}$) & $0.000142$ & $0.00102$ & $0.000304$ & $0.0094$ & $0.00401$ & $0.00577$ & $0.0230$ \\
      CMB+tSZ ($\mathrm{G}+\mathrm{cNG}$) & $0.000144$ & $0.00111$ & $0.000305$ & $0.0121$ & $0.00409$ & $0.00661$ & $0.0347$ \\
      CMB+tSZ ($\mathrm{G}+\mathrm{cNG}+\mathrm{SSC}$) & $0.000144$ & $0.00113$ & $0.000306$ & $0.0130$ & $0.00411$ & $0.00691$ & $0.0376$ \\
    \end{tabular}
  \end{ruledtabular}
  \caption{The expected $1\text{-}\sigma$ error on each parameter
  with CMB anisotropies and tSZ power spectrum.
  We show the values with three different covariances:
  $\mathrm{G}$, $\mathrm{G}+\mathrm{cNG}$, and $\mathrm{G}+\mathrm{cNG}+\mathrm{SSC}$.}
  \label{tab:constraints}
\end{table*}

\section{Conclusions}
\label{sec:conclusions}
The tSZ effect is one of the most important probes in cosmology.
With the power spectrum, we can constrain the cosmological parameters
and investigate the astrophysical effects, e.g., hydrostatic mass bias.
Since any survey is done for a finite volume, even for a full-sky survey,
it is important to realize the impact of
the super-survey modes on the statistical power of tSZ power spectrum.
We quantify the contribution based on the halo model approach
and address biases in parameter estimation.
However, it is found that the super-survey covariance is subdominant
compared with the cNG term, which is sourced
from the trispectrum with the parallelogram configuration.

In order to enhance the statistical significance,
it is proposed that massive nearby galaxy clusters,
which are dominant source of trispectrum,
should be masked to suppress the contribution of the cNG term \cite{Shaw2009}.
We propose the cluster masking based on integrated Compton-$y$ parameter,
which corresponds to the thermal energy content in galaxy clusters,
and the method enables one to significantly reduces the cNG term.
On the other hand, even after masking clusters,
the SSC remains because it originates from relatively less massive clusters
similarly to the power spectrum signal.
Though the overall statistical significance improves,
the SSC has appreciable effect and
weakens the constraints on the parameters.

We have carried out Fisher forecasts on cosmological parameters and
hydrostatic mass bias parameter
through the tSZ power spectrum measurement combining with the results of \textit{Planck}
CMB anisotropy measurements.
In particular, the hydrostatic mass bias parameter,
which cannot be constrained only with primary CMB anisotropies,
can be tightly constrained through the tSZ power spectrum.
By masking clusters, the constraints on the parameters related to
the primordial power spectrum, i.e., $A_\mathrm{s}$ and $n_\mathrm{s}$,
and the hydrostatic mass bias parameter improve.
On the other hand, the SSC becomes important because it persists after masking.
For example, the constraint on the hydrostatic mass bias with covariance including SSC is
$10\%$ larger than that with covariance excluding SSC.
Since the effect of SSC remains at all scales,
it is critical to incorporate the SSC for accurate estimates on parameters
both for ongoing and upcoming observational surveys of the tSZ effect.
Moreover, the cluster masking loses the information from massive galaxy clusters.
In order to compensate for the loss, the joint analysis with cluster counts is thought to be
a practical solution \citep{Oguri2011,Takada2014,Schaan2014}.
The method we developed in this paper increases the potential of tSZ cosmology.
Massive clusters with significant tSZ signals
on individual cluster basis can be used for cosmology,
e.g. via the number counts of the clusters.
On the other hand, when those clusters are masked from the power spectrum measurement,
the tSZ power spectrum can be used to estimate the cosmological parameters.
Thus, we showed that massive clusters and other tSZ clusters
can play complementary roles to cosmology.
This direction is worth further exploring in more detail.


\begin{acknowledgments}
K.O. is supported by JSPS Overseas Research Fellowships.
This work is also supported in part by the World Premier International
Research Center Initiative (WPI Initiative), MEXT, Japan, Reischauer Institute of Japanese Studies at Harvard University, and JSPS
KAKENHI Grant Numbers
JP15H05887, JP15H05893, JP15H05896, JP15K21733, and JP19H00677.
\end{acknowledgments}

\bibliographystyle{apsrev4-2}
\bibliography{main}

\end{document}